\documentclass[pra,preprint,%
author-year,%
superscriptaddress]{revtex4-1}

\usepackage{lineno}

\usepackage[utf8]{inputenc}
\usepackage{graphicx}
\usepackage{dcolumn}
\usepackage{bm}
\usepackage{braket}
\usepackage[utf8]{inputenc}
\usepackage[T1]{fontenc}
\usepackage{mathptmx}
\usepackage{siunitx}
\usepackage{url}
\usepackage{subcaption}
\usepackage{hyperref}
\usepackage{ulem}
\usepackage{siunitx}
\usepackage[export]{adjustbox}

\begin{document}

\title{All Reflective Field-widened Unbalanced Interferometer for Quantum Sensing and Communication Applications}

\author{Ramy Tannous}
\email{ramy.tannous@uwaterloo.ca}
\altaffiliation[Current address:]{ National Research Council of Canada, 100 Sussex Drive, Ottawa ON, Canada}
\affiliation{Institute for Quantum Computing, University of Waterloo, 200 University Avenue W, Waterloo, ON N2L3G1, Canada}
\affiliation{Department of Physics and Astronomy, University of Waterloo, 200 University Avenue W, Waterloo, ON N2L3G1, Canada}

\author{Dogan Sinar}
\affiliation{Institute for Quantum Computing, University of Waterloo, 200 University Avenue W, Waterloo, ON N2L3G1, Canada}
\affiliation{Department of Physics and Astronomy, University of Waterloo, 200 University Avenue W, Waterloo, ON N2L3G1, Canada}

\author{Tabitha D. Arulpragasam}
\altaffiliation[Current address:]{ National Research Council of Canada, 100 Sussex Drive, Ottawa ON, Canada}
\affiliation{Institute for Quantum Computing, University of Waterloo, 200 University Avenue W, Waterloo, ON N2L3G1, Canada}
\affiliation{Department of Chemical Engineering, University of Waterloo, 200 University Avenue W, Waterloo, ON N2L3G1, Canada}

\author{Thomas Jennewein}
\email{thomas.jennewein@uwaterloo.ca}
\affiliation{Institute for Quantum Computing, University of Waterloo, 200 University Avenue W, Waterloo, ON N2L3G1, Canada}
\affiliation{Department of Physics and Astronomy, University of Waterloo, 200 University Avenue W, Waterloo, ON N2L3G1, Canada}
\affiliation{Department of Physics, Simon Fraser University, 8888 University Dr W, Burnaby, BC V5A 1S6, Canada}

\begin{abstract}
Interference of time-bin encoded signals over free-space optical channels typically requires stringent mode filtering on receivers due to wavefront distortions from atmospheric turbulence, conventionally addressed with adaptive optics. Passive multimode receivers based on field-widened interferometers present a compelling alternative, enabling direct interference without the overhead of wavefront correction. We demonstrate a field-widened interferometer design that is implemented solely with reflective surfaces and achieves a high interference visibility (greater than 0.97) for spatially multimode beams. The interference of the multimode beams is enabled by two imaging systems that consist of a cavity configuration between a spherical concave mirror and a flat mirror. The configuration enables small form-factors, is inherently achromatic, and is based on standard spherical mirrors which reduces the complexity of the system. The interferometer is applicable for spatially multimode and turbulent optical channels, such as satellite communication, and is designed for quantum systems that use time-bin encoded qubits. 
\end{abstract}

\maketitle

\section{Introduction}

 Creating time-bin interferometers suitable for free space quantum communication, and quantum sensing applications that do not require adaptive optics have a practical interest~\cite{jin2019genuine,sajeed_observing_2021,tannous2025towards}. These passive interferometers, called field-widened interferometers, reduce the need for adaptive optics by using imaging optics, or carefully chosen refractive indices, to correct angle of incidence induced phase shifts and visibility reductions~\cite{Hilliard:66,hirschberg_field_1974,mahadevan_inexpensive_2008}, thus reducing the required technical overhead which is important for platforms that have limited resources, such as drones and satellites. Field-widened interferometers have been investigated as time-bin interferometers that can be used with spatially multimode channels~\cite{jin2019genuine, vallone2016interference,zeitler2016super,jin2018demonstration,tannous2023fully,wu2024single}. Interferometers that have a large physical footprint are highly susceptible to thermal and vibrational noise, making them impractical beyond highly controlled environments. More recent efforts reported improvements in the thermal stability of the interferometers \cite{cahall_multi-mode_2020}. However, interferometers that employ optically dense materials (refractive interferometers) to create the corrective imaging system will experience chromatic dispersion, which limits their usability with many quantum light sources, which are relatively broadband~\cite{anwar2021entangled}. More recently, some analysis on the visibility of standard unbalanced Michelson interferometers with multimode signals was carried out~\cite{tretiakov2024multi}. However, such an approach, though simple, is limited to larger spot sizes, shorter interferometer path delays, and a small field-of-view. The latter is the most significant because small deviations from a normal incident beam cause significant drops in the interference visibility, which is problematic for alignment and quantum imaging applications. Here, we report a field-widened interferometer design that only uses reflective optics that have limited chromatic dependence, and a cavity imaging system that enables long path delays with smaller physical footprints. The interferometer uses a configuration similar to an Offner relay~\cite{monson_bircam_2009}, so we refer to it as the Offner relay interferometer (ORI). By taking advantage of broadband operating range of reflective optics and the cavity imaging system, the ORI's design is a step towards robust, field-deployable quantum interferometers.

\section{Optical Design}
\begin{figure}[htb]
	\centering

    \begin{subfigure}[b]{0.38\textwidth}
         \centering
         \includegraphics[width=\textwidth]{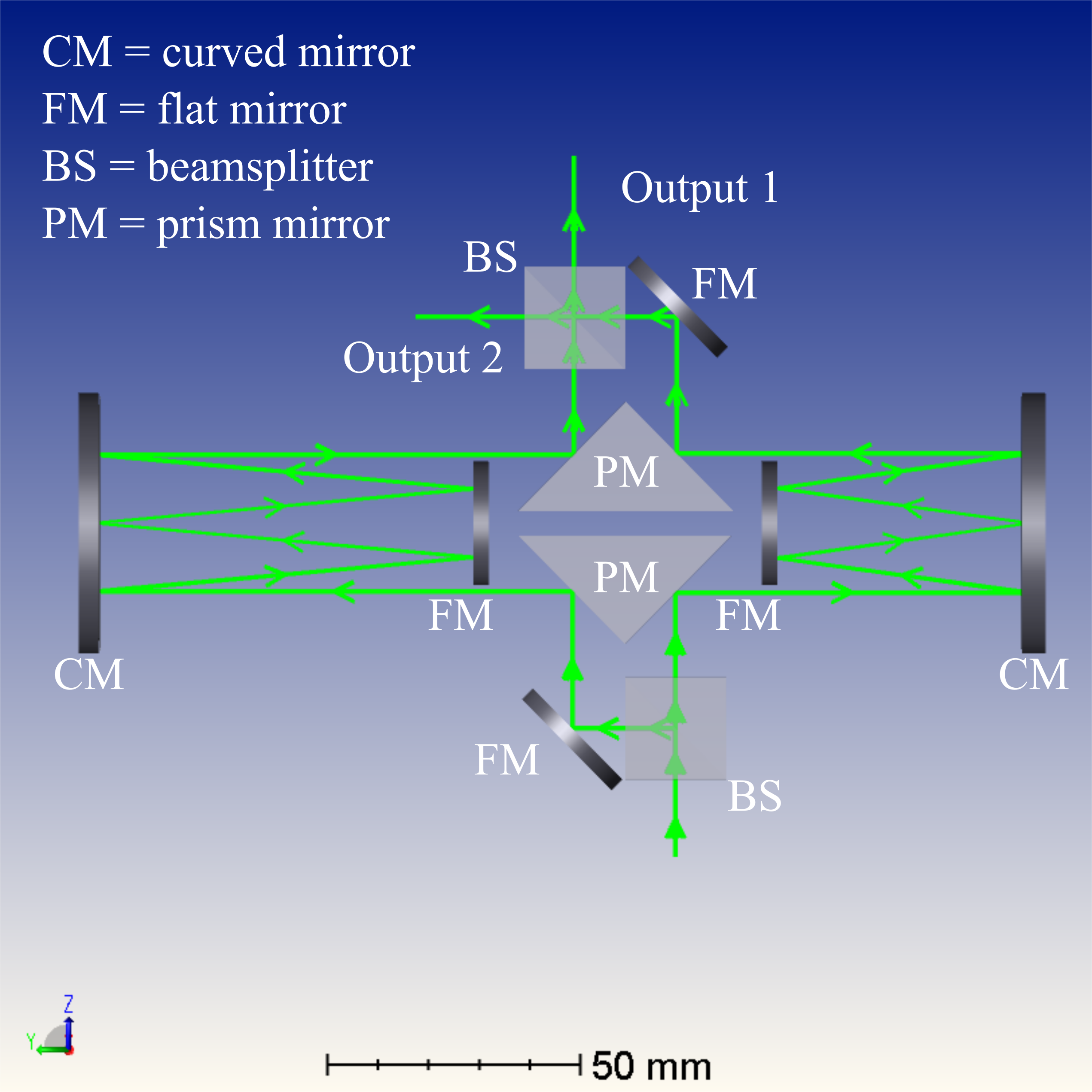}
         \caption{\SI{500}{ps}}
         \label{fig:ORI-top}
    \end{subfigure}
    \begin{subfigure}[b]{0.38\textwidth}
         \centering
         \includegraphics[width=\textwidth]{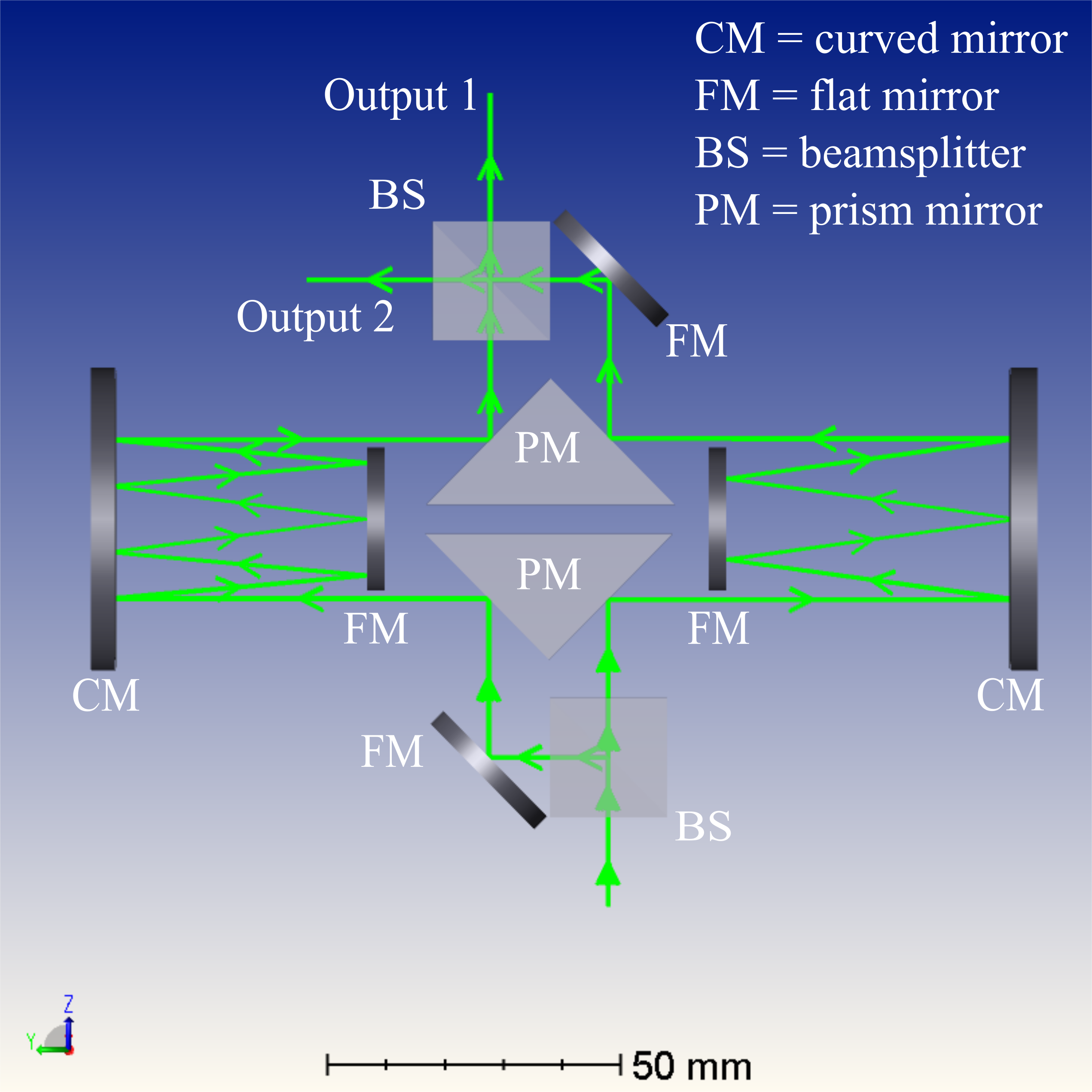}
         \caption{\SI{715}{ps}}
         \label{fig:smaller-ORI-top}
    \end{subfigure}
       \begin{subfigure}[b]{0.38\textwidth}
         \centering
         \includegraphics[width=0.8\textwidth]{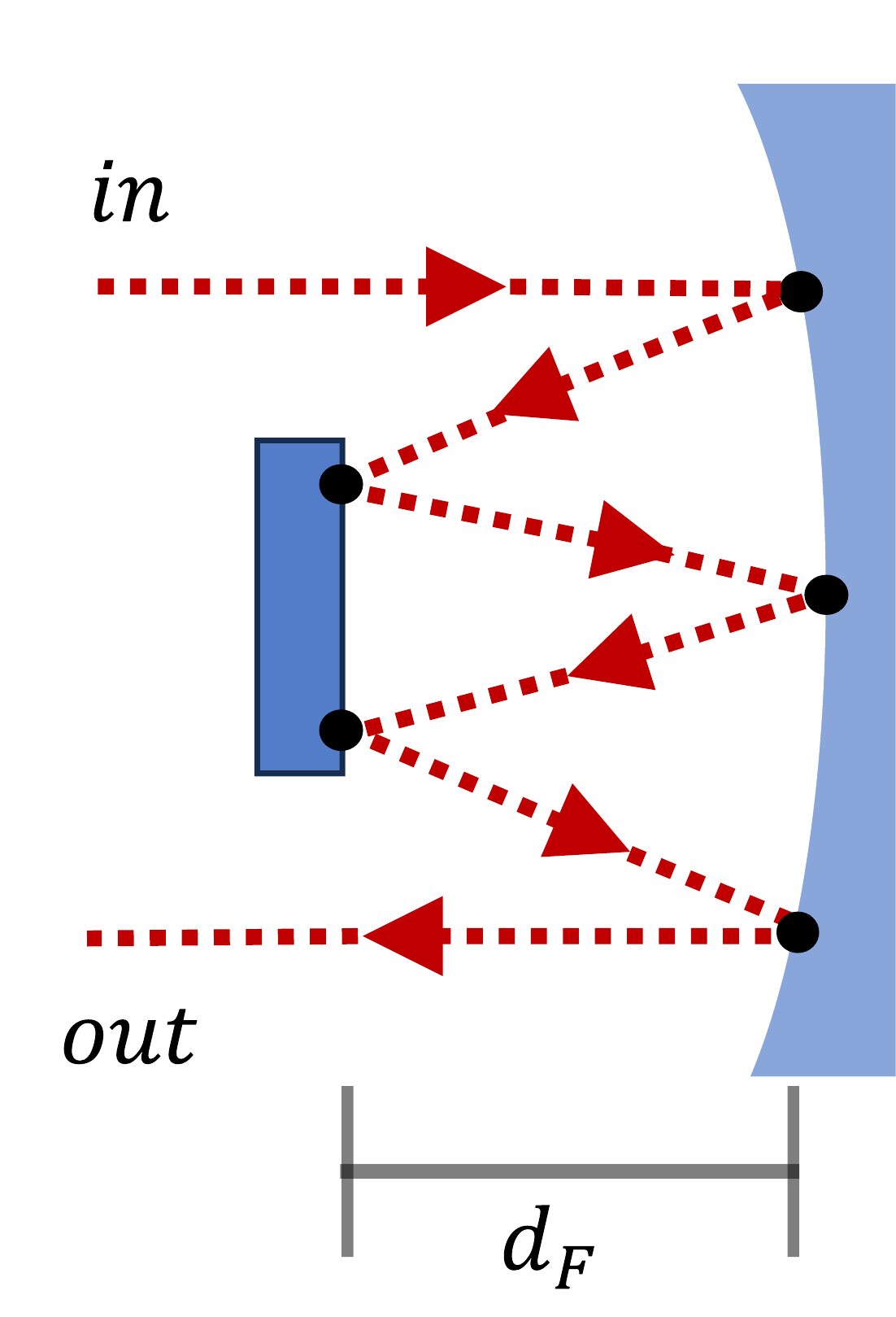}
         \caption{}
         \label{fig:d_f}
    \end{subfigure}
    
	  \caption{Schematic of the optical design of the Offner relay interferometer.  (\subref{fig:ORI-top}) Model of an ORI with a \SI{500}{ps} relative path delay. (\subref{fig:smaller-ORI-top}) Increased relative path delay (\SI{715}{ps}) by adjusting $d_f$ to create additional reflections in the folded configuration while reducing physical footprint of the system. The radius of curvature of the CM are kept constant in (\subref{fig:ORI-top}) and (\subref{fig:smaller-ORI-top}). (\subref{fig:d_f}) $d_f$ is the critical distance of the imaging system and folded configuration of the ORI.}
	    \label{fig:setup}
	\end{figure}

 The optical design of the ORI is shown in Fig.~\ref{fig:setup} and is driven by the following objectives. The first is a small physical footprint that is enabled by using a folded optical path in the imaging system that consists of a spherical mirror and a flat mirror in a cavity configuration. This allows the relative delay between the two arms of the interferometer to be longer than the physical footprint. The folded optical path also permits the use of long focal length spherical mirrors, decreasing the spherical aberration in the system. In contrast, an actual Offner relay~\cite{monson_bircam_2009} uses two spherical mirrors, which does not easily allow for multiple reflections in a cavity configuration. The second is to use off-the-shelf, simple optical components. Spherical curved mirrors are widely available, and have the advantage of simpler alignment compared to parabolic mirrors. The third is to minimize chromatic aberrations, and dispersion inherent to refractive interferometers. The ORI avoids using refractive optics and instead uses reflective optics for the imaging system, therefore can be used over a broad wavelength range.
    
 The folded optical path is created by adjusting the distance $d_f$ (Fig.~\ref{fig:d_f}) between the spherical mirror (CM) and the flat mirror (FM) of the imaging system. Table.~\ref{tab:ORI-dimensions} compares the ORI designs with interferometer systems that use optically dense materials and refractive optics. As shown in Table.~\ref{tab:ORI-dimensions}, the folded Offner relay can achieve smaller physical footprints, especially for designs with larger time delays. All values in Table.~\ref{tab:ORI-dimensions} are obtained using 1-inch standard optical components except for the CM which has a 2-inch aperture. The overall relative delay between the two paths of the ORI depends on the CM focal length, and can be changed by adjusting $d_f$, altering the number of reflections on the spherical mirror, $B$. Fig.~\ref{fig:ORI-top} and~\ref{fig:smaller-ORI-top} show two ORI configurations with the same radius of curvature for CM while $d_f$ is reduced.

    \begin{table}[htbp]
        \caption{Various design configurations and simulation results of ORI using \SI{25}{mm} aperture optics and CM aperture of \SI{50}{mm}. The form factors include an estimated length and width of the optical setup, height is not considered. The form factor required for a Michelson interferometer (MI) that employs N-BK10 and N-SF66, having the same relative path delay, and optics with \SI{25}{mm} aperture size is compared. $R$ is the radius of curvature of CM and $B$ is the number of reflections on the CM. The visibilities reported here are found using numerical simulations.}
        \centering
        \label{tab:ORI-dimensions}
        {
        \begin{tabular*}{\linewidth}{@{\extracolsep{\fill}}cccccc@{}}
        \hline
        $R_1$~/~$R_2$ [mm]  & $B_1$~/~$B_2$ & Time Delay [ps] & ORI Vis SM~/~MM &  ORI Size [mm$^2$] & MI Size [mm$^2$] \\ \hline
        200~/~300        & 3~/~3               & 501         & 0.9995~/ 0.9994  & 5460 & 5195 \\ 
        200~/~300        & 4~/~3               & 715         & 0.9990~/ 0.9985  & 4953 & 7185 \\
        150~/~300        & 3~/~3               & 768         & 0.9993~/ 0.9993  & 5080 & 7847 \\
        150~/~300        & 4~/~3               & 896         & 0.9971~/ 0.9964  & 4830 & 8870 \\
        150~/~300        & 5~/~3               & 1020        & 0.9903~/ 0.9892  & 4572 & 10021\\
        \hline 
        \end{tabular*}
        }
    \end{table}
    
\section{Optical Performance}
    \subsection{Simulations}
\begin{figure}[htbp]
		\centering
		\begin{subfigure}[t]{0.4\textwidth}
			\centering
			\includegraphics[width=\textwidth]{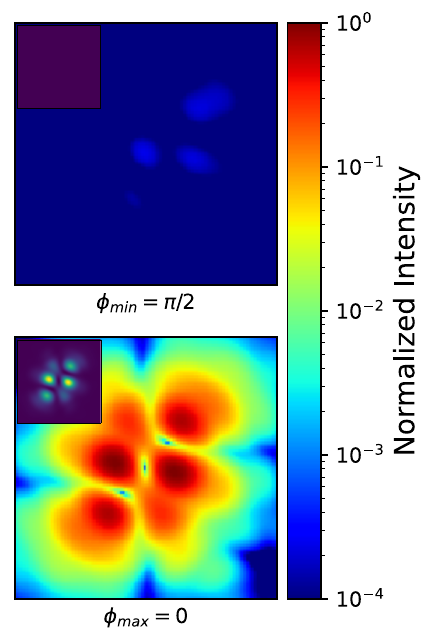}
			\caption{Normal incidence}
			\label{fig:spot_overlap_p25}
		\end{subfigure}
		\begin{subfigure}[t]{0.4\textwidth}
			\centering
			\includegraphics[width=\textwidth]{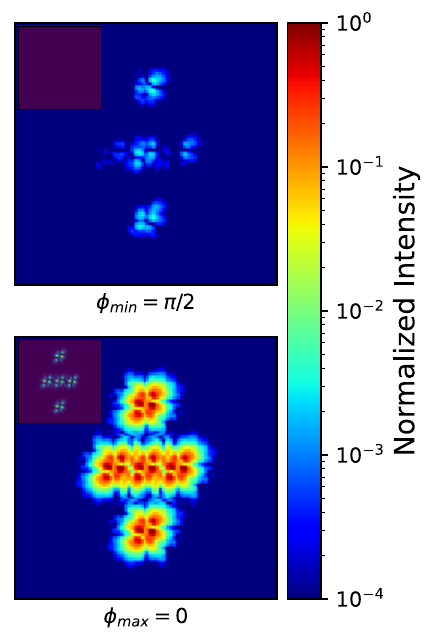}
			\caption{Angular displacement}
			\label{fig:spot_overlap_p7}
		\end{subfigure}
		
		\caption{ Simulated coherent interference maxima (bottom) and minima (top) using non-sequential ray tracing. (\subref{fig:spot_overlap_p25}) A single higher order input field with normal incidence. (\subref{fig:spot_overlap_p7}) Multiple input fields with angular offsets of \SI{0.3}{\degree} in the horizontal direction and \SI{0.58}{\degree} in the vertical direction. The calculated visibility across the entire detector aperture is (\subref{fig:spot_overlap_p25}) $0.9994$ (\subref{fig:spot_overlap_p7}) $0.9939$. The insets show the intensity in linear scaling.}
		\label{fig:spot_overlap_tog}
	\end{figure}

    To demonstrate that the ORI functions as a field-widened interferometer, i.e. a multimode interferometer, we verified the design using numerical optical ray tracing simulations. The results of the simulations are shown in Table.~\ref{tab:ORI-dimensions} for several configurations. In particular, Fig.~\ref{fig:spot_overlap_tog} discusses in more detail an ORI with $R_1=200$~mm, $R_2=300$~mm, $B_1=3$, and $B_2=3$, and with a CM aperture of \SI{50}{\milli\meter}. We quantify the interference quality using the Michelson visibility defined as,
    \begin{equation}
        \label{eq:vis}
        V=\frac{I_{\textit{max}}-I_{\textit{min}}}{I_{\textit{max}}+I_{\textit{min}}},
    \end{equation}
	where $I_{\textit{max(min)}}$ are the maximum (minimum) intensities of the interference fringes. The input mode in the numerical simulation is a pure Gaussian and a higher order Hermite Gauss (HG) mode. The sum of the output port intensity across the entire detection surface in Fig.~\ref{fig:spot_overlap_p25} top and bottom is used to calculate the interference, in this case $0.9994$. Table.~\ref{tab:ORI-dimensions} indicates that the high visibility results are consistent for all the ORI configurations for both a pure Gaussian and a higher order HG mode entering on-axis and with normal incidence. 
    
    In addition to the analysis at normal incidence, the input signals are given small angular offsets of up to \SI{0.3}{\degree} in the horizontal direction, and \SI{0.58}{\degree} in the vertical direction. Despite having different angles of incidence, the outputs of the five beams all interfere in phase with each other, as shown in Fig.~\ref{fig:spot_overlap_p7} with a visibility of $0.9939$. The horizontal offset is limited by beam clipping when entering and exiting the folded optical path with the FM acting as the limiting optic, which can be adjusted by using different optical element sizes. The vertical direction is limited by a drop in visibility as the angular offset increases. However, for the particular design used for these simulations, a maximal angle of \SI{1.72}{\degree} is achieved before a 4\% drop in the interference visibility. This maximum input angle varies with the ORI design parameters.
        
\begin{figure}[htbp]
        \centering
        
        \includegraphics[width=0.75\columnwidth]{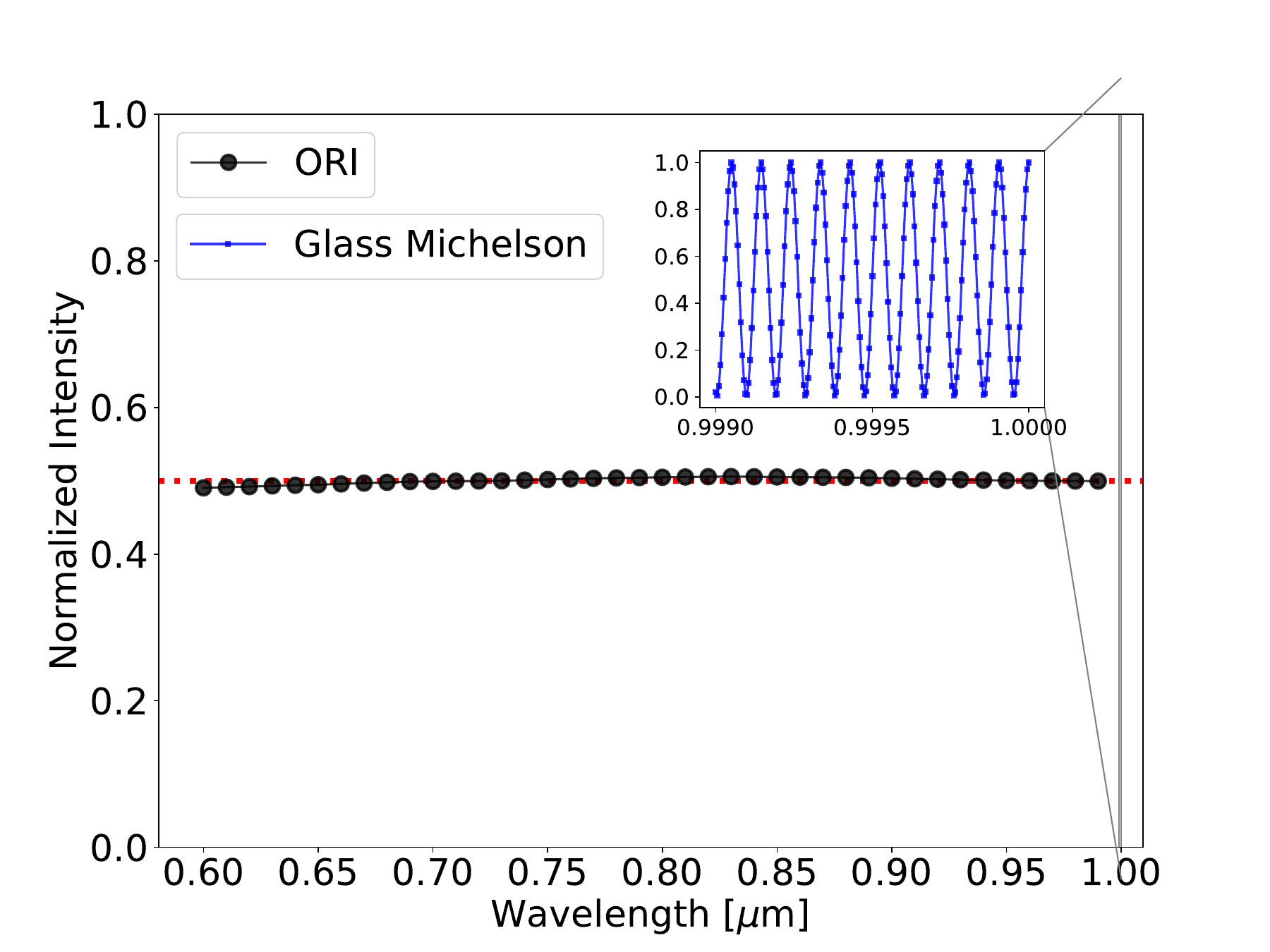}
        \caption{The main figure shows the intensity of one of the outputs of the ORI as a function of wavelength. For comparison, the inset shows the intensity of the output of a field-widened glass Michelson interferometer over a smaller wavelength range. The red dotted line is at $0.5$ for examination.}
        \label{fig:white-light-sim}
    \end{figure}

   Another practical advantage of the ORI is its broad operating range that is enabled by the use of an all-reflective imaging system. Thus making the ORI effectively achromatic within the range of the beamsplitter's coatings. We confirm the achromatic capabilities by numerical simulations for a broadband light source. The simulated pulsed light is passed through an unbalanced Michelson then through the ORI. The results in Fig.~\ref{fig:white-light-sim} indicate that despite the wide wavelength range of the signal, the intensity at the output of the ORI is relatively flat as a function of wavelength when compared to the dotted red line. Therefore, indicating that the total relative phase of the combined Michelson and ORI simulation setup is wavelength independent across the large range of \SIrange{600}{1000}{\nano\meter}. The normalized intensity in Fig.~\ref{fig:white-light-sim} is at $0.5$ since the relative phase of the combined Michelson and ORI simulation setup is set to $\pi/4$ where the system should be the most sensitive to any wavelength dependent phase changes. In contrast to the ORI, the inset of Fig.~\ref{fig:white-light-sim} shows the relative phase of the output of a refractive field-widened interferometer that uses N-BK10 (chromatic dispersion: \SI{-0.019469}{\per\micro\meter} at \SI{785}{\nano\meter}) and N-SF66 (chromatic dispersion: \SI{-0.085037}{\per\micro\meter} at \SI{785}{\nano\meter}) has a strong chromatic dependence due to the material dispersion. Thus refractive interferometers can be difficult to use with broadband sources of light.

    \subsection{Experimental Optical performance}

	A prototype ORI was built using parameters that align with those used to create Fig.~\ref{fig:spot_overlap_tog}, (i.e. $R_1, R_2=200, 300$~mm and $B_1, B_2=3, 3$). The visibility of the ORI is experimentally measured using both single mode and multimode signals from a \SI{15}{\kilo\hertz} bandwidth continuous wave \SI{785}{nm} external cavity diode laser (Toptica DL pro). For the single mode signals, the measured visibility is $0.977\pm0.016$ represents the visibility maximum that the interferometer can achieve. Note that with single mode signals such high visibilities are easily achievable with uncorrected standard interferometers such as a Michelson interferometer, which in comparison showed a single mode visibility of $0.997\pm0.022$. 
	
\begin{figure}[htbp]
    \centering
    \begin{subfigure}[t]{0.45\linewidth}
			\centering
			\includegraphics[width=\linewidth]{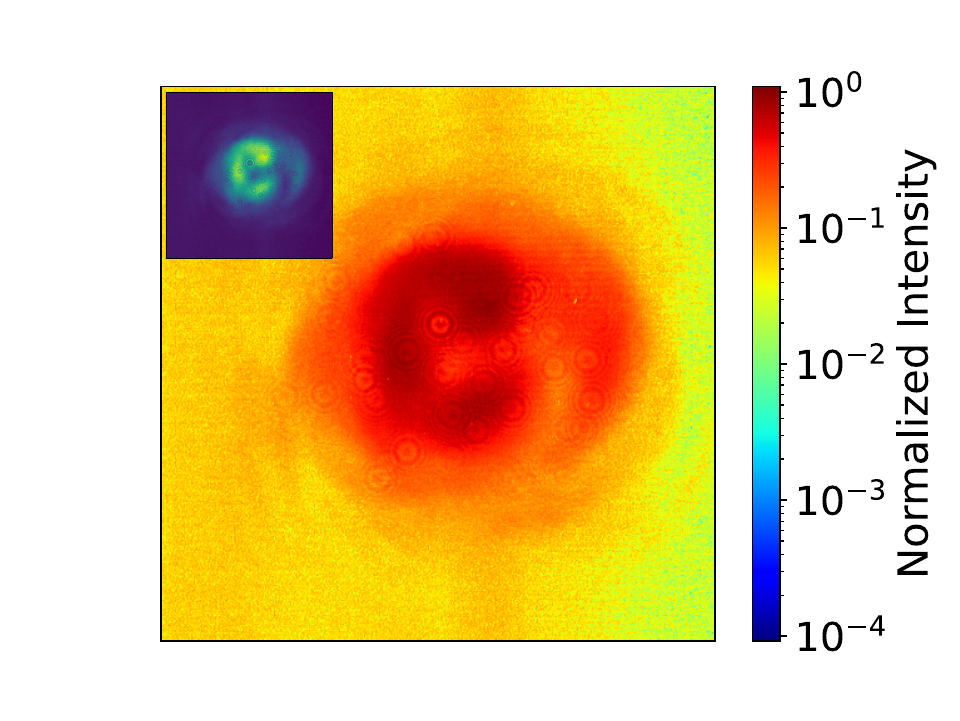}
			\caption{Constructive interference}
			\label{fig:mmf_bright}
		\end{subfigure}
		\begin{subfigure}[t]{0.45\linewidth}
			\centering
			\includegraphics[width=\linewidth]{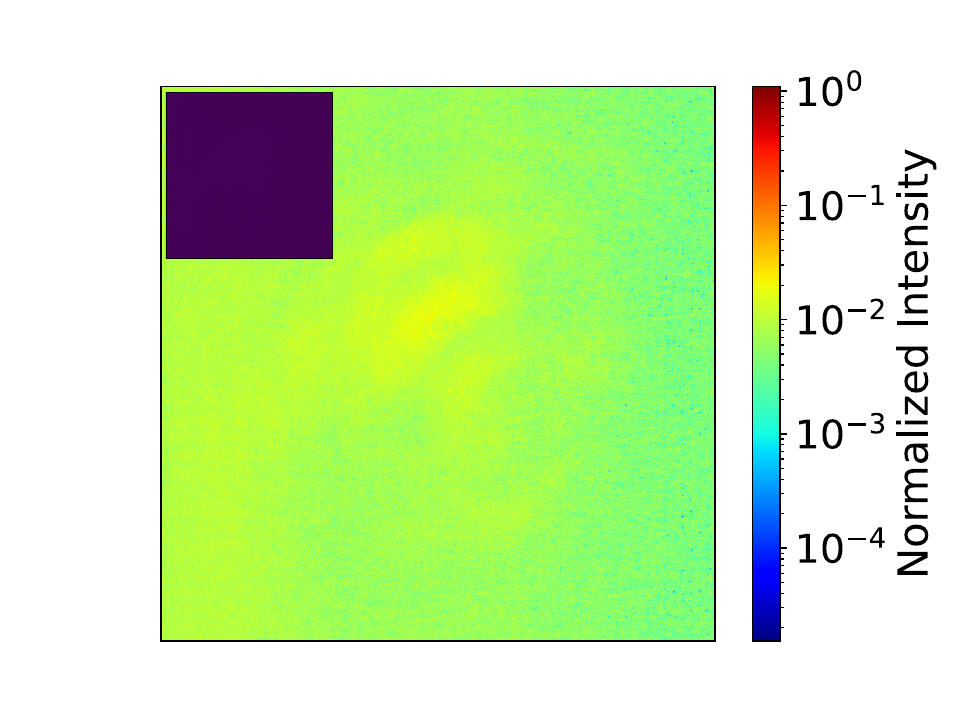}
			\caption{Destructive interference}
			\label{fig:mmf_dark}
		\end{subfigure}
    \begin{subfigure}[t]{0.75\columnwidth}
        \centering
        \includegraphics[width=\textwidth]{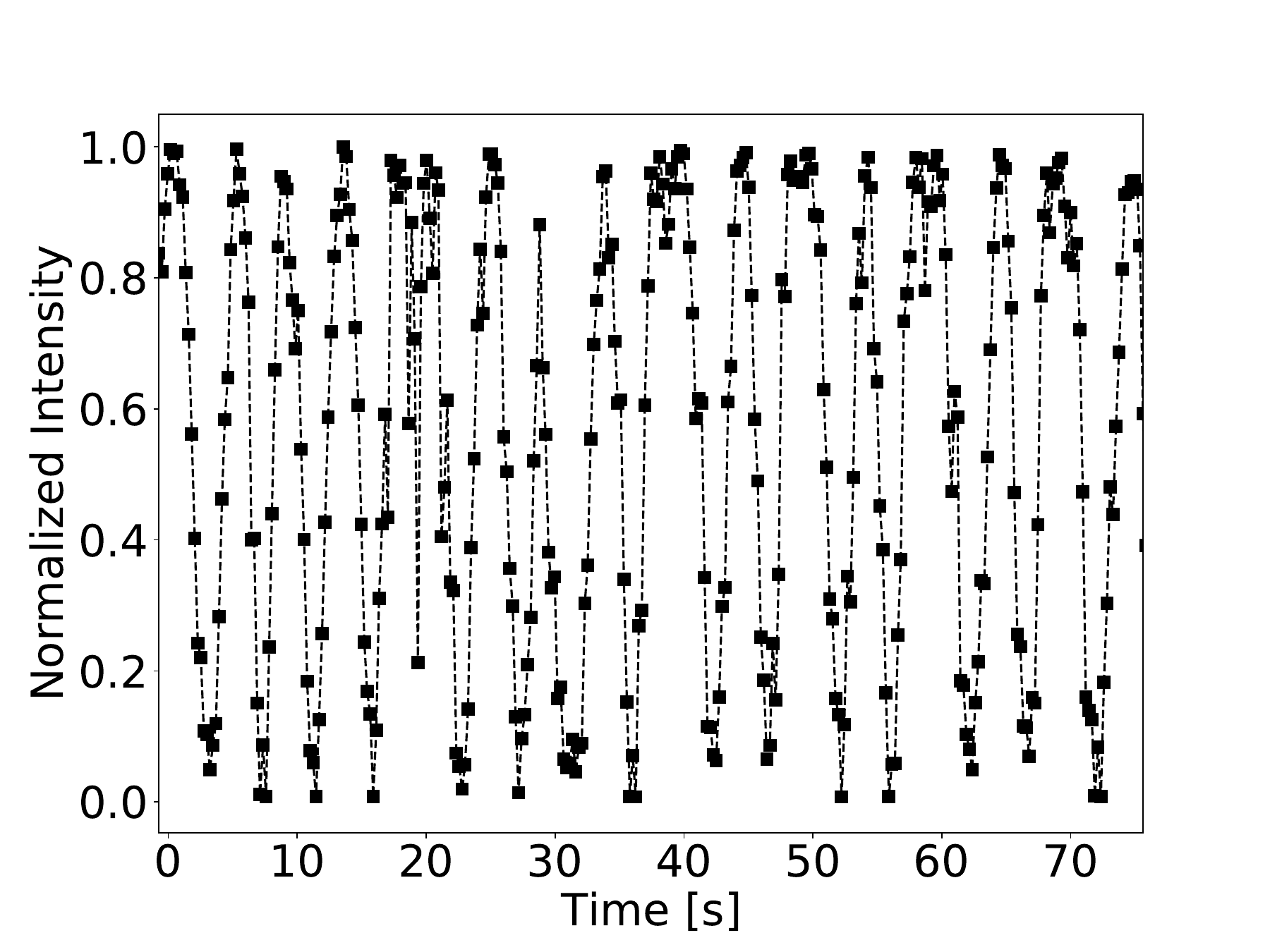}    
        \caption{}
        \label{fig:ex-vis-graph}
    \end{subfigure}

     \caption{Normalized multimode signal intensity measured at the output of the ORI with a camera. (\subref{fig:mmf_bright}) Constructive interference and (\subref{fig:mmf_dark}) destructive interference. (\subref{fig:ex-vis-graph}) Sinusoidal intensity fluctuations from phase variations induced by a piezoelectric actuator. Measurement error are represented by the size of the plot markers. The intensity is normalized such that the maximum recorded value in the constructive interference is $1$ and the detector background is $0$. The insets show the intensity in linear scaling.}
     \label{fig:Offner-vis}
\end{figure}

	The true benefit of a field-widened interferometer is revealed when it is used with multimode signals. The visibility of the multimode signal may be lower than that for the single mode case due to many factors that cause deviations from an ideal imaging system, i.e. optical aberrations. The optical design consistently achieved visibilities of $0.9$ or greater with little to no alignment effort. However, the remaining increase to the theoretical maximum required careful alignment of the CM-FM cavity. We measured a maximum interference visibility of the ORI with multimode signals to be $0.979\pm0.022$, shown in Fig.~\ref{fig:Offner-vis}. Fig.~\ref{fig:mmf_bright} and~\ref{fig:mmf_dark} show the normalized intensity for constructive and destructive interference, respectively, as imaged by a camera device (WinCamD-UCD12). Overall, the multimode visibility of the ORI greatly exceeded that of an regular unbalanced Michelson interferometer which had a visibility of $0.536\pm0.023$ with the same multimode signals and spot size. Furthermore, we compared the ORI performance to the field-widened Mach Zehnder interferometers used in \cite{jin2019genuine} that had a measured visibility for a multimode signal of $0.970\pm0.013$. Thus, the performance of the ORI is comparable to other field-widened interferometers however, as shown in Table.~\ref{tab:ORI-dimensions} the ORI has the potential for smaller form factors while being non-dispersive.

\subsection{Imaging demonstration}

\begin{figure}[htbp]
    \centering
    \includegraphics[width=0.75\textwidth]{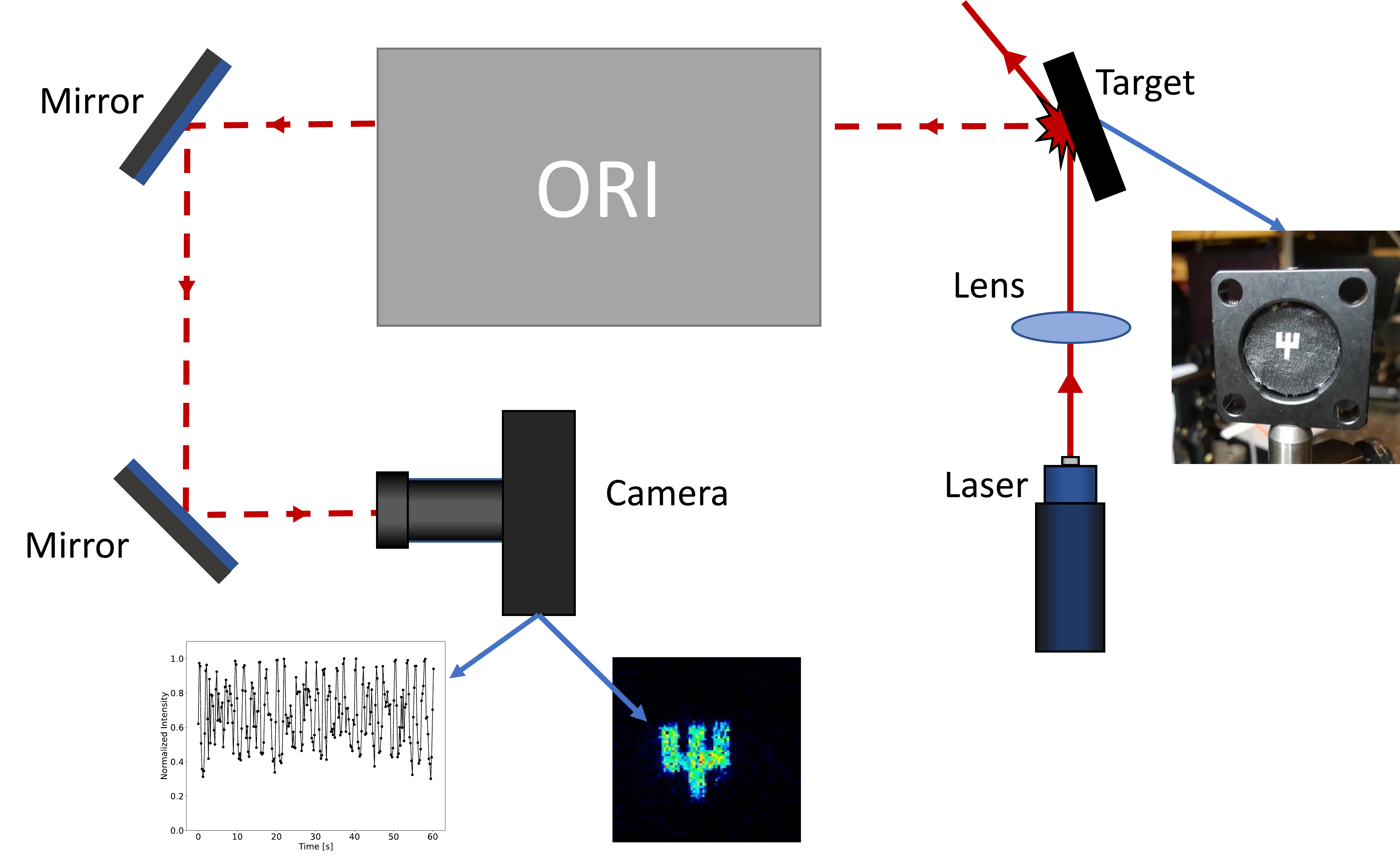}
    \caption{Quantum sensing demonstration using the ORI system. The signal is diffusely scattered off the target and directed through the two paths of the ORI. The relative path difference of the ORI is being changed using a piezoelectric actuator which allows for coherence properties of the signals to be measured.}
    \label{fig:prac-image}
\end{figure}
\begin{figure}[htbp]
    \centering
    \includegraphics[width=0.75\linewidth]{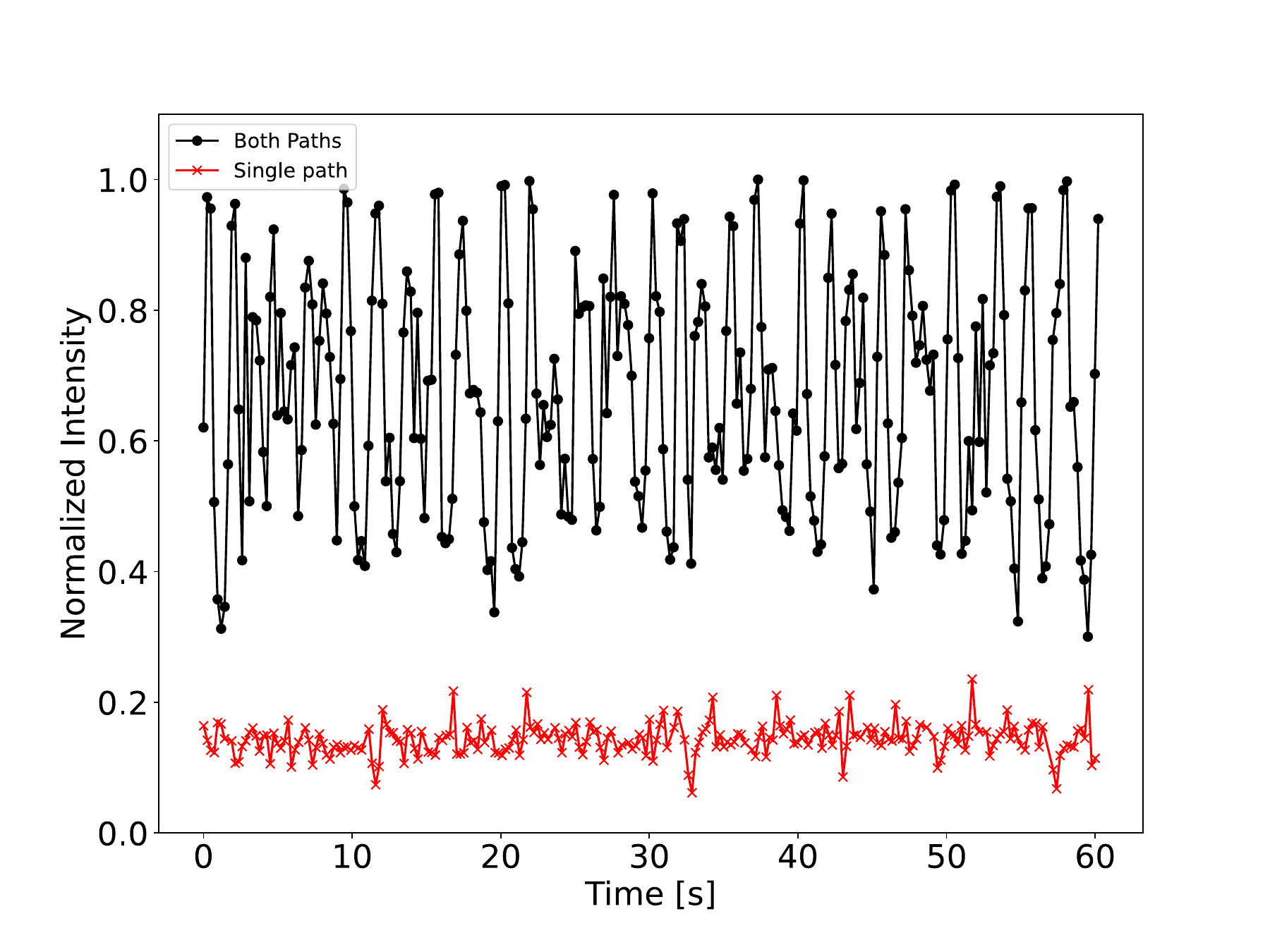}
    \caption{Intensity of the psi target as imaged through the ORI. The black circles are the intensity pattern of the scattered light from the target when allowed to pass through both arms of the interferometer. The red crosses show the stable intensity when only one arm is used, and the other arm is blocked. The intensity is normalized to the maximum value of the data set of the black circles. The red crosses are lowered by $0.3$ to increase the clarity of the plot. }
    \label{fig:res-image-1}
\end{figure}

We further demonstrate the imaging capabilities of the ORI by observing an object through the ORI that is scattering light from a \SI{785}{\nano\meter} laser, shown in Fig.~\ref{fig:prac-image}. Both the image of the target and the interference of the signals are measured simultaneously. The results of the demonstration are found in Fig.~\ref{fig:res-image-1} which indicate that the ORI can be used for imaging interferometry. The interferometer system combined with a camera is able to image the target while simultaneously measuring the phase information of the signals, thus showing its usability in quantum enhanced LiDAR~\cite{sajeed_observing_2021}. This point is further stressed in Figures~\ref{fig:res-image-1} where the intensity of the signal after passing through the ORI are shown for two scenarios: passing through both ORI paths, and blocking one ORI path. In both scenarios a piezoelectric actuator is being driven to produce phase modulation which will cause intensity fringes. The visibility is found to be around $0.49 \pm 0.03$ for the full interferometer scenario, compared to $0.11\pm0.04$ observed when blocking one path of the interferometer. The different scenarios demonstrate that the observed interference visibility with both arms unblocked is not an artifact, whereas the blocked case has no periodic interference despite the presence of the piezoelectric actuator. The relatively reduced visibility in the imaging scenario can be attributed to poor alignment of the system, high background light, and rapid phase fluctuations over the integration time. Regardless, the results indicate that the ORI is a field-widened interferometer as it demonstrated interference of scattered light.

\section{Conclusion}

In summary, we present a novel optical design for a field-widened time-bin interferometer that can easily integrate with current free space and fiber optical networks. The interferometer design uses reflective optics for the imaging system which allows for direct use with multiple single photon sources at different wavelength ranges, and for spectral multiplexing. We experimentally demonstrated that the reflective optical design performs well as an interferometer with a multimode interference visibility of greater than $0.97$. Furthermore, the folded optical paths of the ORI design is capable of producing long relative path delays with a significantly reduced form factor compared to a standard field-widened Michelson approach. Future work is to improve the ease of alignment by reducing the number of optical elements involved. Although our investigation in this manuscript is limited to the configurations shown in Table.~\ref{tab:ORI-dimensions}, different spherical mirror radii and optic sizes can be used to increase the relative path delay while simultaneously reducing the physical footprint of the ORI. The reduced form-factor enables the ability to use alternative manufacturing techniques to house the optical setup, such as additive manufacturing~\cite{TannousRamy2023}. Future work will investigate the limits of the design in terms of optical relative path delay and form factor, and work towards testing the device in a practical field setting.

\section{Acknowledgments}
We thank Dr. Shihan Sajeed and Dr. Youn Seok Lee for helpful discussion. RT would like to thank the National Science and Engineering Research Council of Canada, CGS-D for personal funding. This project was funded by the Deference Research and Development Canada IDEAS program.

	\bibliography{references}

@article{anwar2021entangled,
  title={Entangled photon-pair sources based on three-wave mixing in bulk crystals},
  author={Anwar, Ali and Perumangatt, Chithrabhanu and Steinlechner, Fabian and Jennewein, Thomas and Ling, Alexander},
  journal={Review of Scientific Instruments},
  volume={92},
  number={4},
  year={2021},
  publisher={AIP Publishing}
}

@article{sajeed_observing_2021,
  title={Observing quantum coherence from photons scattered in free-space},
  author={Sajeed, Shihan and Jennewein, Thomas},
  journal={Light: Science \& Applications},
  volume={10},
  number={1},
  pages={121},
  year={2021},
  publisher={Nature Publishing Group UK London}
}

@article{cahall_multi-mode_2020,
	title = {Multi-mode {Time}-delay {Interferometer} for {Free}-space {Quantum} {Communication}},
	volume = {13},
	issn = {2331-7019},
	doi = {10.1103/PhysRevApplied.13.024047},
	language = {en},
	number = {2},
	urldate = {2020-11-29},
	journal = {Physical Review Applied},
	author = {Cahall, Clinton and Islam, Nurul T. and Gauthier, Daniel J. and Kim, Jungsang},
	month = feb,
	year = {2020},
	pages = {024047},
}

@article{jin2019genuine,
  title={Genuine time-bin-encoded quantum key distribution over a turbulent depolarizing free-space channel},
  author={Jin, Jeongwan and Bourgoin, Jean-Philippe and Tannous, Ramy and Agne, Sascha and Pugh, Christopher J and Kuntz, Katanya B and Higgins, Brendon L and Jennewein, Thomas},
  journal={Optics express},
  volume={27},
  number={26},
  pages={37214--37223},
  year={2019},
  publisher={Optical Society of America}
}

@article{jin2018demonstration,
  title={Demonstration of analyzers for multimode photonic time-bin qubits},
  author={Jin, Jeongwan and Agne, Sascha and Bourgoin, Jean-Philippe and Zhang, Yanbao and L{\"u}tkenhaus, Norbert and Jennewein, Thomas},
  journal={Physical Review A},
  volume={97},
  number={4},
  pages={043847},
  year={2018},
  publisher={APS}
}

@article{monson_bircam_2009,
	title = {{BIRCAM}: {A} {Near}-{Infrared} {Camera} for {The} {University} of {Wyoming} {Red} {Buttes} {Observatory}},
	volume = {121},
	issn = {0004-6280, 1538-3873},
	shorttitle = {{BIRCAM}},
	url = {http://iopscience.iop.org/article/10.1086/603619},
	doi = {10.1086/603619},
	language = {en},
	number = {881},
	urldate = {2021-12-05},
	journal = {Publications of the Astronomical Society of the Pacific},
	author = {Monson, Andrew J. and Pierce, Michael J.},
	month = jul,
	year = {2009},
	pages = {728--734},
}

@article{hirschberg_field_1974,
	title = {Field {Widened} {Michelson} {Spectrometer} with {No} {Moving} {Parts}},
	volume = {13},
	issn = {0003-6935, 1539-4522},
	url = {https://www.osapublishing.org/abstract.cfm?URI=ao-13-2-233},
	doi = {10.1364/AO.13.000233},
	language = {en},
	number = {2},
	urldate = {2022-01-20},
	journal = {Applied Optics},
	author = {Hirschberg, Joseph G.},
	month = feb,
	year = {1974},
	pages = {233},
}

@article{mahadevan_inexpensive_2008,
  title={An inexpensive field-widened monolithic Michelson interferometer for precision radial velocity measurements},
  author={Mahadevan, Suvrath and Ge, Jian and Fleming, Scott W and Wan, Xiaoke and DeWitt, Curtis and Van Eyken, Julian C and McDavitt, Dan},
  journal={Publications of the Astronomical Society of the Pacific},
  volume={120},
  number={871},
  pages={1001},
  year={2008},
  publisher={IOP Publishing}
}

@inproceedings{zeitler2016super,
  title={Super-dense teleportation for space applications},
  author={Zeitler, Chris and Graham, Trent M and Chapman, Joseph and Bernstein, Herbert and Kwiat, Paul G},
  booktitle={Free-Space Laser Communication and Atmospheric Propagation XXVIII},
  volume={9739},
  pages={312--316},
  year={2016},
  organization={SPIE}
}

@article{vallone2016interference,
  title={Interference at the single photon level along satellite-ground channels},
  author={Vallone, Giuseppe and Dequal, Daniele and Tomasin, Marco and Vedovato, Francesco and Schiavon, Matteo and Luceri, Vincenza and Bianco, Giuseppe and Villoresi, Paolo},
  journal={Physical review letters},
  volume={116},
  number={25},
  pages={253601},
  year={2016},
  publisher={APS}
}

@article{tannous2023fully,
  title={Towards Fully Passive Time-Bin Quantum Key Distribution over Multi-Mode Channels},
  author={Tannous, Ramy and Wu, Wilson and Vinet, St{\'e}phane and Perumangatt, Chithrabhanu and Sinar, Dogan and Ling, Alexander and Jennewein, Thomas},
  journal={arXiv preprint arXiv:2302.05038},
  year={2023}
}

@phdthesis{TannousRamy2023,
author={{Tannous, Ramy}},
title={Advancing the robustness of polarization and time bin quantum key distribution for free-space channels},
year={2023},
publisher="UWSpace",
school={University of Waterloo},
url={http://hdl.handle.net/10012/19444}
}

@article{wu2024single,
  title={Single-Photon Interference over 8.4 km Urban Atmosphere: Toward Testing Quantum Effects in Curved Spacetime with Photons},
  author={Wu, Hui-Nan and Li, Yu-Huai and Li, Bo and You, Xiang and Liu, Run-Ze and Ren, Ji-Gang and Yin, Juan and Lu, Chao-Yang and Cao, Yuan and Peng, Cheng-Zhi and others},
  journal={Physical Review Letters},
  volume={133},
  number={2},
  pages={020201},
  year={2024},
  publisher={APS}
}

@article{tretiakov2024multi,
  title={A multi-mode free-space delay interferometer with no refractive compensation elements for phase-encoded QKD protocols},
  author={Tretiakov, VV and Kravtsov, KS and Klimov, AN and Kulik, SP},
  journal={Laser Physics Letters},
  volume={21},
  number={6},
  pages={065206},
  year={2024},
  publisher={IOP Publishing}
}

@article{Hilliard:66,
author = {R. L. Hilliard and G. G. Shepherd},
journal = {J. Opt. Soc. Am.},
number = {3},
pages = {362--369},
publisher = {Optica Publishing Group},
title = {Wide-Angle Michelson Interferometer for Measuring Doppler Line Widths$\ast$},
volume = {56},
month = {Mar},
year = {1966},
url = {https://opg.optica.org/abstract.cfm?URI=josa-56-3-362},
doi = {10.1364/JOSA.56.000362},
}

@article{tannous2025towards,
  title={Towards fully passive time-bin quantum key distribution over moving free-space channels},
  author={Tannous, Ramy and Wu, Wilson and Vinet, St{\'e}phane and Perumangatt, Chithrabhanu and Sinar, Dogan and Ling, Alexander and Jennewein, Thomas},
  journal={Optics Express},
  volume={33},
  number={17},
  pages={35635--35648},
  year={2025},
  publisher={Optica Publishing Group}
}
	
\end{document}